\documentclass{PoS}

\title{TeV blazars and their distance}

\ShortTitle{TeV blazars and their distance}

\author{\speaker{Elisa Prandini}\\% \thanks{A footnote may follow.}\\
       Padova University \& INFN\\
       E-mail: \email{prandini@pd.infn.it}}
\author{Giacomo Bonnoli\\
        Osservatorio di Merate\\
        E-mail: \email{giacomo.bonnoli@brera.inaf.it}}
\author{Laura Maraschi\\
        Osservatorio di Brera\\
        E-mail: \email{laura.maraschi@brera.inaf.it}}
\author{Mos\'e Mariotti\\
        Padova University \& INFN\\
        E-mail: \email{mose.mariotti@pd.infn.it}}
\author{Fabrizio Tavecchio\\
        Osservatorio di Merate\\
        E-mail: \email{fabrizio.tavecchio@brera.inaf.it}}
\abstract{Recently, a new method to constrain the distance of blazars with unknown redshift using combined observations in the GeV and TeV regimes has been developed,  with the underlying assumption that the Very High Energy (VHE) spectrum corrected for the absorption of TeV photons by the Extragalactic Background Light (EBL) via photon-photon interaction should still be softer than the gamma-ray spectrum observed by Fermi/LAT. The constraints found  are related to the real redshifts by a simple linear relation, that has been used to infer the unknown distance of blazars. The sample will be revised with the up-to-date spectra in both TeV and GeV bands, the method tested with the more recent EBL models and finally applied to the unknown distance blazars detected at VHE.}

\FullConference{Cosmic Radiation Fields: Sources in the early Universe\\
                November 9-12, 2010\\
                Desy, Germany}

\begin{document}

\section{Introduction}
The extragalactic TeV sky catalogue ($E>100$ GeV), counts nowadays 45 objects\footnote{For an
updated list see: http://www.mppmu.mpg.de/$\sim$rwagner/sources/}. 
Many of these sources have been recently detected also at
GeV energies by the {\it Fermi} satellite~\cite{abdo10,abdo09}, allowing 
for the first time a quasi-continuous coverage of the spectral shape 
of extragalactic VHE emitters over more than 4 decades of energy. 
The large majority of extragalactic TeV emitting objects
are blazars, radio-loud active galactic nuclei with a 
relativistic jet closely oriented towards  the Earth, as described in \cite{urry}.
The photon flux emitted by a blazar in both GeV and TeV regimes
can be usually well approximated with power laws,
of the form $dN/dE = f_0 (E/E_0)^{-\Gamma}$, where $\Gamma$
is the power law index.
An important effect involving VHE photons emitted by cosmological sources 
is the production of electron-positron pairs 
($\gamma \gamma \rightarrow e^{+}e^{-}$), caused by the interaction 
with the so-called Extragalactic Background Light (EBL) \cite{stecker92}.
Quantitatively, the effect of the interaction of VHE photons with EBL
is an exponential attenuation of the flux 
by a factor $\tau (E,z)$, where $\tau$
is the optical depth, function of 
both photon energy and source redshift. Thus, the observed
differential energy spectrum from a blazar 
is related to the emitted one according 
to $F_{\rm obs}(E)=e^{-\tau(E)} F_{\rm em}(E)$.
EBL is composed of stellar light emitted and partially 
reprocessed by dust throughout the entire history of 
cosmic evolution. 
Due to the lack of direct EBL knowledge, many models
have been elaborated in the last
years~\cite{dominguez10,franceschini08,kneiske10,stecker06}, but
the uncertainties remains quite large.
Moreover, for some blazars, the derivation of the
intrinsic spectrum is difficult due to the uncertainty or lack of a
 measurement of the redshift. 

In a recent paper, we have proposed a method to derive an estimate on 
the distance of a blazar \cite{prandini10}.
The method is based on the comparison between the spectral index at GeV energies
as measured by LAT after 5.5 months of data taking,  basically 
unaffected by the cosmological absorption, 
and the TeV spectrum corrected for the absorption. 
In that work, it is  shown that according to present observations,
the spectral slope measured by LAT, 
$\Gamma_{LAT}$, in the energy range 0.2--300\,GeV, 
can be considered as a limiting slope for the 
emitted spectrum at TeV energies (i.e. corrected for EBL absorption). 
This maximum hardness hypothesis was successfully 
tested on a sample of 14 well--known distance sources.
Consequently, the redshift, $z^*$, at which 
the deabsorbed TeV slope equals $\Gamma_{LAT}$, 
can be used as an upper limit for the source distance.
An empirical relation between the
upper limit, $z^*$, and the true redshift of a blazar was then found. 
A simple linear relation fits well the $z^*$--$z_{true}$ distribution, 
for three different EBL models. 
The relation is  associated to the
linear expression, found in \cite{stecker10}, for the steepening of the observed 
TeV slopes due to EBL absorption.
Hence, $z^*$ and $z_{true}$ 
are related by a linear function of the form $z^*=A+Bz_{true}$.
This relation can be used to 
give an estimate on the source distance.

In this paper we present an update of that work, based on the more
recent LAT catalogue \cite{abdo10}. We test the validity of the
maximum hardness hypothesis and that of the linear relation
between $z^*$ and $z_{true}$, using a new EBL model \cite{dominguez10}.
A cosmological scenario with $h=0.72$, $\Omega_M=0.3$ 
and $\Omega_\Lambda=0.7$ is assumed.

\section{Analysis and results}
\begin{center}
   \begin{table*}
     {\small
       \centering
       \begin{tabular}{llcccc}
         \hline
         \hline
         Source Name     & $z_{true}$  & $\Gamma_{LAT}$ &  $\Gamma_{LAT}$ & $z^*_{Fra}$ & $z^*_{Dom} $\\
                         &            & (5.5\,m)  & (1\,y)  &      &   \\
        \hline
        \hline
        Mkn 421          & 0.030 & 1.78 $\pm$ 0.03  & 1.81 $\pm$ 0.02  &   0.07 $\pm$ 0.02  &  0.07 $\pm$ 0.02  \\
        Mkn 501          & 0.034 & 1.73 $\pm$ 0.06  & 1.85 $\pm$ 0.04  &   0.08 $\pm$ 0.02  &  0.07 $\pm$ 0.02   \\
        1ES 2344$+$514   & 0.044 & 1.76 $\pm$ 0.27  & 1.57 $\pm$ 0.17  &   0.19 $\pm$ 0.03  &  0.18 $\pm$ 0.03  \\
        Mkn 180          & 0.045 & 1.91 $\pm$ 0.18  & 1.86 $\pm$ 0.11  &   0.21 $\pm$ 0.11  &  0.20 $\pm$ 0.11  \\
        1ES 1959$+$650   & 0.047 & 1.99 $\pm$ 0.09  & 2.09 $\pm$ 0.05 &    0.07 $\pm$ 0.03  & 0.07 $\pm$ 0.03   \\
        BL Lacertae      & 0.069 & 2.43 $\pm$ 0.10  & 2.37 $\pm$ 0.04 &    0.27 $\pm$ 0.14  & 0.26 $\pm$ 0.15\\
        PKS 2005$-$489   & 0.071 & 1.91 $\pm$ 0.09  & 1.90 $\pm$ 0.06 &    0.18 $\pm$ 0.03  & 0.18 $\pm$ 0.03 \\
        W Comae          & 0.102 & 2.02 $\pm$ 0.06  & 2.06 $\pm$ 0.04 &    0.24 $\pm$ 0.05  & 0.23 $\pm$ 0.05 \\
        PKS 2155$-$304   & 0.116 & 1.87 $\pm$ 0.03  & 1.91 $\pm$ 0.02 &    0.22 $\pm$ 0.01  & 0.21 $\pm$ 0.01 \\
        RGB J0710$+$591  & 0.125 & 1.21 $\pm$ 0.25  & 1.28 $\pm$ 0.21 &    0.21 $\pm$ 0.06  & 0.20 $\pm$ 0.06 \\
        1ES 0806$+$524   & 0.138 & 2.04 $\pm$ 0.14  & 2.09 $\pm$ 0.10 &    0.23 $\pm$ 0.15  & 0.22 $\pm$ 0.15 \\
        H 2356$-$309     & 0.165 & -                & 2.10 $\pm$ 0.17 &    0.16 $\pm$ 0.07  & 0.16 $\pm$ 0.07 \\
        1ES 1218$+$304   & 0.182 & 1.63 $\pm$ 0.12  & 1.70 $\pm$ 0.08 &    0.21 $\pm$ 0.08  & 0.20 $\pm$ 0.08 \\
        1ES 1101$-$232   & 0.186 & -                & 1.36 $\pm$ 0.58 &    0.23 $\pm$ 0.11  & 0.22 $\pm$ 0.10 \\
        1ES 1011$+$496   & 0.212 & 1.82 $\pm$ 0.05  & 1.93 $\pm$ 0.04 &    0.60 $\pm$ 0.18  & 0.60 $\pm$ 0.18 \\
        S5 0716$+$714    & 0.310$^a$ & 2.16 $\pm$ 0.04  & 2.15 $\pm$ 0.03 &    0.23 $\pm$ 0.10  & 0.22 $\pm$ 0.10 \\
        PG 1553+113      & 0.400 & 1.69 $\pm$ 0.04 &  1.66 $\pm$ 0.03 &    0.75 $\pm$ 0.07  & 0.75 $\pm$ 0.07 \\
        3C~66A           & 0.444$^a$ & 1.93 $\pm$ 0.04  & 1.92 $\pm$ 0.02 &    0.39 $\pm$ 0.05  & 0.38 $\pm$ 0.05 \\
        3C~279           & 0.536 & 2.34 $\pm$ 0.03  & 2.32 $\pm$ 0.02 &   -   &  - \\
        \hline
        \hline
      \end{tabular}
      \caption
       {List of TeV blazars used in this study. Source name (first column), $\Gamma_{LAT}$ reported in the 5.5 months catalogue (second column) and in the first year catalogue (third column), $z^*$ value, as described in the text, obtained with the Franceschini model (fifth column) and the Dominguez model (sixth column). $^a$: uncertain. \label{table_values}}
    }
  \end{table*}
\end{center}
The updated sample, presented in this study,
 is composed by all the extragalactic TeV 
emitters located at redshift larger than $z=0.01$ and 
detected by LAT after the first year of data taking \cite{abdo10}.
In total, there are 16 sources with well known redshift and two
additional sources with uncertain redshift, namely 3C~66A and S5~0716+714. 
Despite many TeV emitters have been discovered since late 2009 by
Cherenkov Telescopes, none of the new sources could
be included in our sample, since the corresponding
TeV spectra are not published, yet.
In the first column of Table~\ref{table_values}, we list the 
sources used in the study. The second and third columns represent 
the slopes measured by LAT after 5.5~months (between 0.2--300\,GeV) 
and one year (between 0.1--100\,GeV).
Three new sources are included in the sample: 
RGB~J0710$+$591, H~2356$-$309 and 1ES~1101$-$232, located at redshifts 
0.125, 0.165 and 0.186, respectively. 
The last two sources were not detected by LAT in the first 5.5 months, 
while the spectrum of RGB~J0710$+$591 has only
recently been published by the VERITAS collaboration \cite{acciari10b}. 
With respect to the 5.5 months catalogue, the new LAT determination of the
spectral slopes is characterized by smaller errors, 
due to the increased statistics. 

With this enlarged  data set, we estimate the quantity $z^*$, redshift at which the deabsorbed
TeV spectrum exhibits the same slope measured by LAT at lower energies. 
We adopt the mean energy density EBL model \cite{franceschini08}, 
hereafter Franceschini model\footnote{The absorption values
used here are taken directly from the WEB site http://www.astro.unipd.it/background,
and differ slightly  from those used in \cite{prandini10}, where an 
extrapolation method was used.}, and the
new model \cite{dominguez10}, hereafter Dominguez model. 
The values obtained are listed in the last
two columns of Table~\ref{table_values}.
In case of 3C~279, the slopes of deabsorbed spectrum do not converge to the 
LAT value for any redshift.
  \begin{figure}
    \centering  
    \includegraphics[width=4.in]{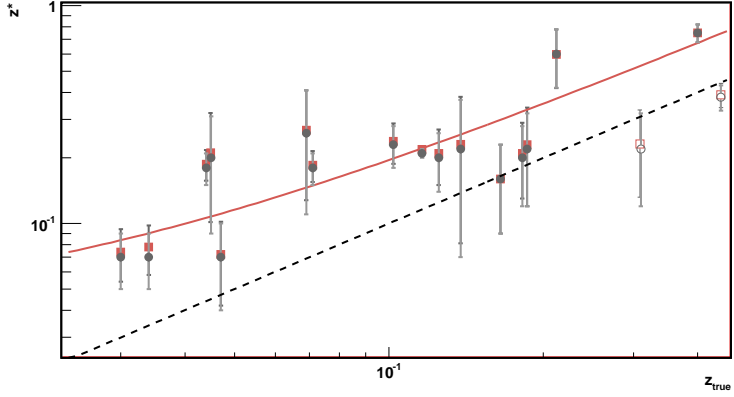}
    \caption{$z^*$ versus true redshift derived with the 
      procedure described in the text,
      for two EBL models (Franceschini and Dominguez, squares and 
      circles respectively). The open symbols represents the two sources with
      uncertain redshift, 3C~66A and S5~0716+714, not used in 
      the fits. The dashed line is the bisector, 
      while the continuous line is the 
      linear fit to the data obtained using the Franceschini model.}
    \label{correlationplot}
  \end{figure} 

Figure~\ref{correlationplot} represents the distribution $z_{true}~-~z^*$, 
obtained with the two EBL models. 
All the $z^*$ values distribute above or on the bisector.
This confirms that  $z^*$ can be considered 
an upper limit on the source redshift, hence the 
maximum hardness hypothesis is confirmed also in this study.
The linear curve drawn represents the fit of the data derived with 
the Franceschini model. The linear trend of the distribution
is less evident here than in the previous study. 
The probability of the fit is, in fact, 
$\sim$6\%, well below the previous value (58\%).
The reason for this behaviour can be related 
to the new sources introduced, but also to the 
new LAT determination of the slopes. 
In order to investigate such alternatives, we have fitted the 
distribution excluding the
three new sources. The new fit  returns a probability of 
9\%, close to the value obtained with the full sample.
This result suggests that the low probability found is mainly due to the
smaller error bars characterizing the determination of the new slopes in 
the GeV band with respect to previous estimates.
The parameters found with the Dominguez model, Table~\ref{table:fit_parameters},
are similar to those obtained with the Franceschini model. 

\begin{table}
  \centering
  \begin{tabular}{ l c c}
  \hline
  \hline
  EBL Model    & $A$ & $B$    \\
  \hline
  \hline
 Franceschini   & 0.036 $\pm$ 0.014  & 1.60 $\pm$ 0.14  \\
 Dominguez      & 0.030 $\pm$ 0.014 & 1.58 $\pm$ 0.14 \\ 
  \hline
  \hline
\end{tabular}
\caption{Parameters of the linear fitting curves ($z^*$~=~$A+Bz_{true}$), 
  obtained with the two EBL models. \label{table:fit_parameters}}
\end{table}

\begin{figure*}
    \centering
    \includegraphics[width=0.4\textwidth]{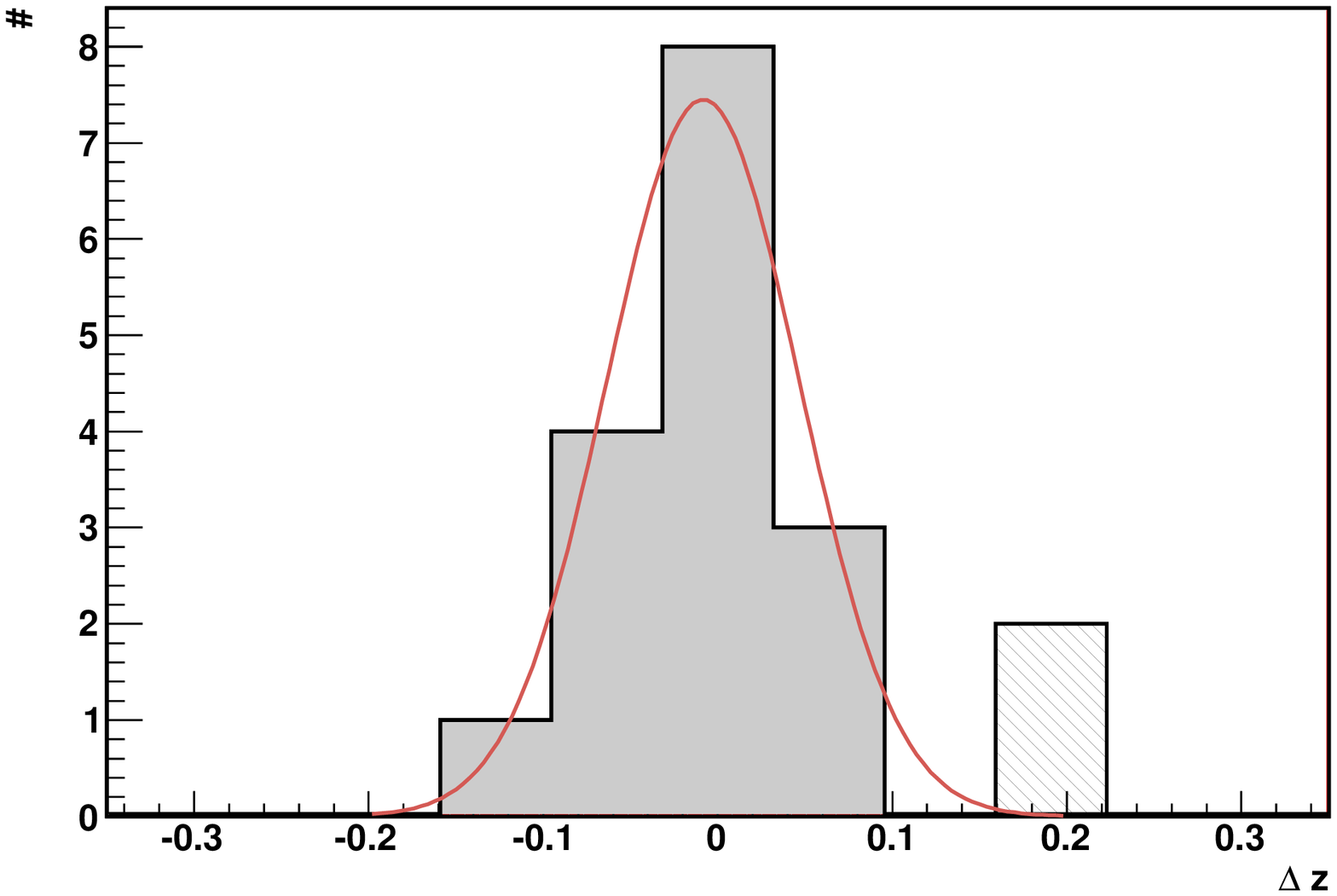}
    \includegraphics[width=0.4\textwidth]{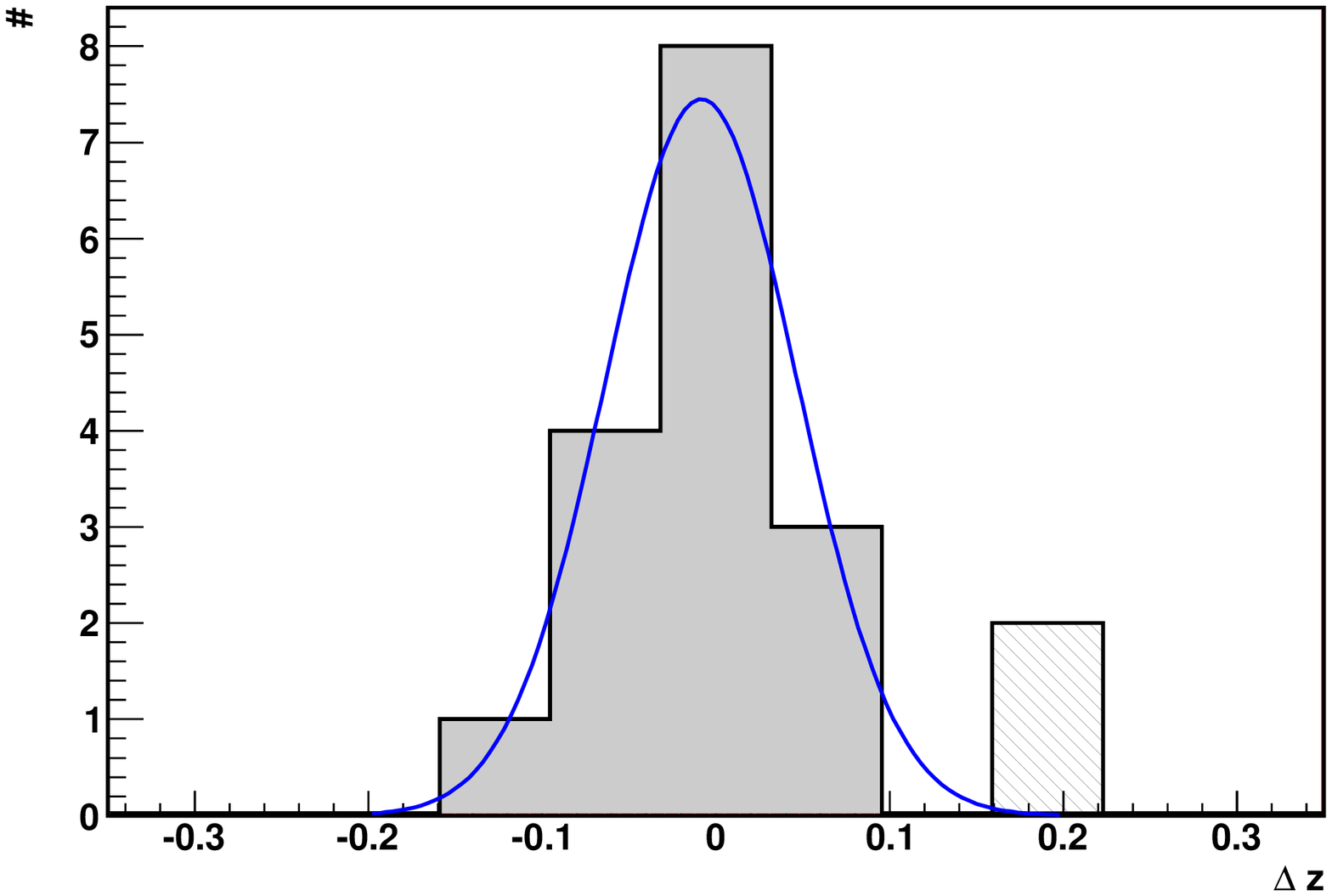}
    \caption{Dispersion $\Delta z$ ($z_{true} - z_{rec}$), obtained with the 
      Franceschini model (left plot) and the Dominguez model (right plot).
      The shaded area represent the two sources with 
      uncertain redshift (S5~0716+714 and 3C~66A), not used in the Gaussian fits.}
    \label{fig:dispersion_fradom_fermi1y}
  \end{figure*}

Following the first study, 
we investigate the distribution $\Delta z$, difference between
the values $z_{rec}$, listed in Table~\ref{table:linearz_zrec_updatedsample}, obtained
by inverting the linear formula $z_{rec}$~=~($z^*-A)/B$, and the true redshifts, $z_{true}$. 
The histograms obtained with the two EBL models, Figure~\ref{fig:dispersion_fradom_fermi1y},
are well fitted by a Gaussian of $\sigma~=~0.05$, which can be assumed as the error on the 
reconstructed redshift, $z_{rec}$, estimated with this method. 
In both histograms, the two sources with uncertain redshift, not used
for the Gaussian fits, lie outside the expected interval. 
This result confirms that the behaviour of S5~0716+714 and 3C~66A
is different from that found for other sources, suggesting that or 
these sources are peculiar, or their redshift is wrong.

In conclusion, we can say that with an enlarged data set 
the results previously found are confirmed. However, the 
linearity of the $z^*$--$z_{true}$ relation has a smaller 
probability, due to the reduced 
errors of the new $\Gamma_{LAT}$ determinations.

 \begin{center}
   \begin{table*}
     {\small
       \centering
       \begin{tabular}{llcccc}
         \hline
         \hline
         Source Name     & $z_{true}$  & $z_{rec(Fra)}$ & $z_{rec(Dom)} $\\
        \hline
        \hline
        Mkn 421          & 0.030 &   0.02 & 0.02  \\
        Mkn 501          & 0.034 &   0.03 & 0.02  \\
        1ES 2344$+$514   & 0.044 &   0.09 & 0.09  \\
        Mkn 180          & 0.045 &   0.11 & 0.11  \\
        1ES 1959$+$650   & 0.047 &   0.02 & 0.02  \\
        BL Lacertae      & 0.069 &   0.14 & 0.14  \\
        PKS 2005$-$489   & 0.071 &   0.09 & 0.09  \\
        W Comae          & 0.102 &   0.13 & 0.13  \\
        PKS 2155$-$304   & 0.116 &   0.11 & 0.11  \\
        RGB J0710$+$591  & 0.125 &   0.11 & 0.11  \\
        1ES 0806$+$524   & 0.138 &   0.12 & 0.12  \\
        H 2356$-$309     & 0.165 &   0.08 & 0.08  \\
        1ES 1218$+$304   & 0.182 &   0.11 & 0.11  \\
        1ES 1101$-$232   & 0.186 &   0.12 & 0.12  \\
        1ES 1011$+$496   & 0.212 &   0.35 & 0.36  \\
        S5 0716$+$714    & 0.310 &   0.12 & 0.12  \\
        PG 1553+113      & 0.400 &   0.45 & 0.45  \\
        3C~66A           & 0.444 &   0.22 & 0.22 \\
        \hline
        \hline
      \end{tabular}
      \caption
       {Reconstructed redshift, $z_{rec}$, with the background models \cite{franceschini08,dominguez10}. 
         The error on the values, determined a posteriori, is 0.05 in both cases. See text for details.}
      \label{table:linearz_zrec_updatedsample}
    }
  \end{table*}
\end{center}

\section{The redshift of PKS~1424+240}
\begin{figure}
     \centering
     \includegraphics[width=0.45\textwidth]{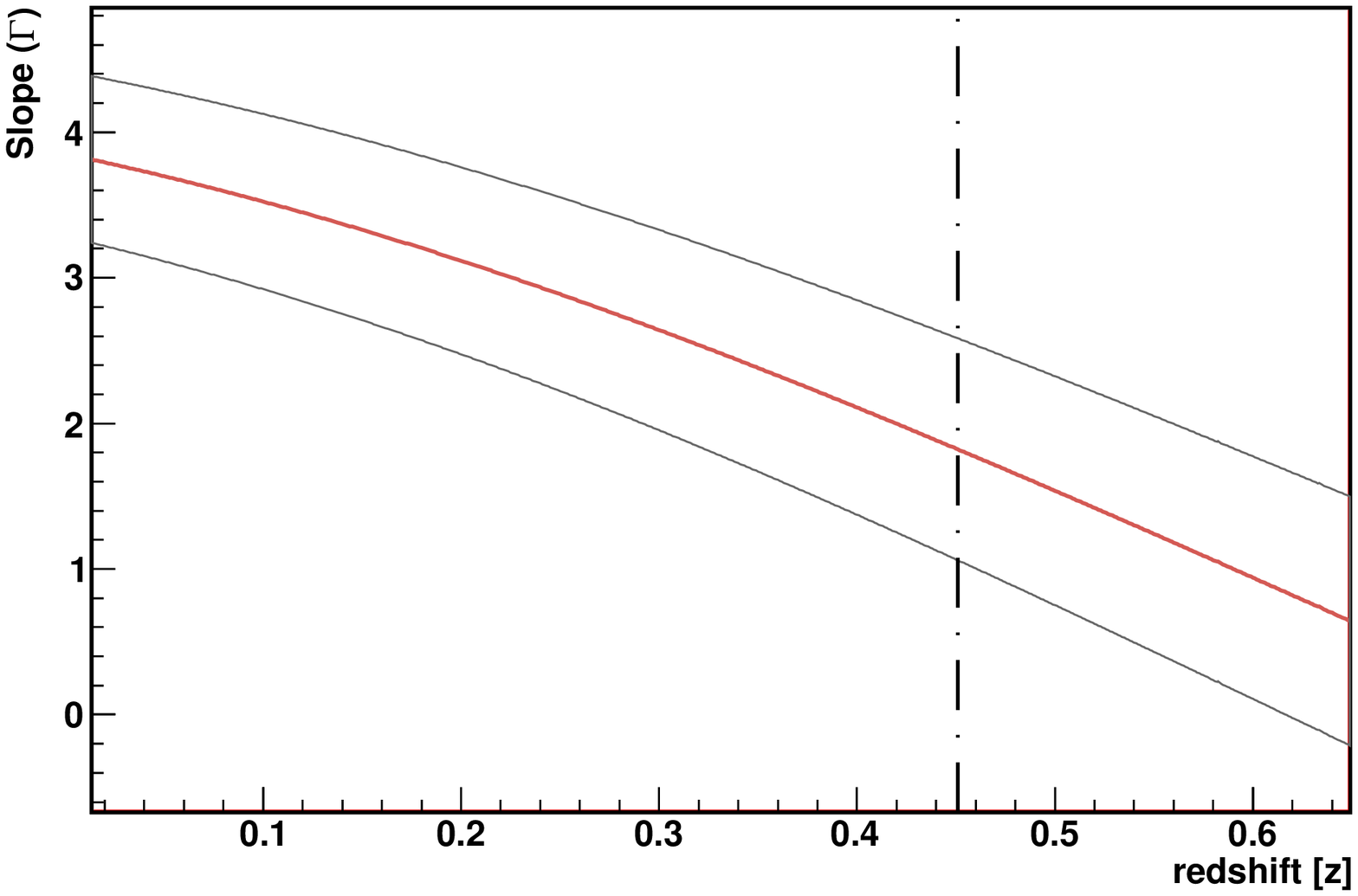}
     \includegraphics[width=0.45\textwidth]{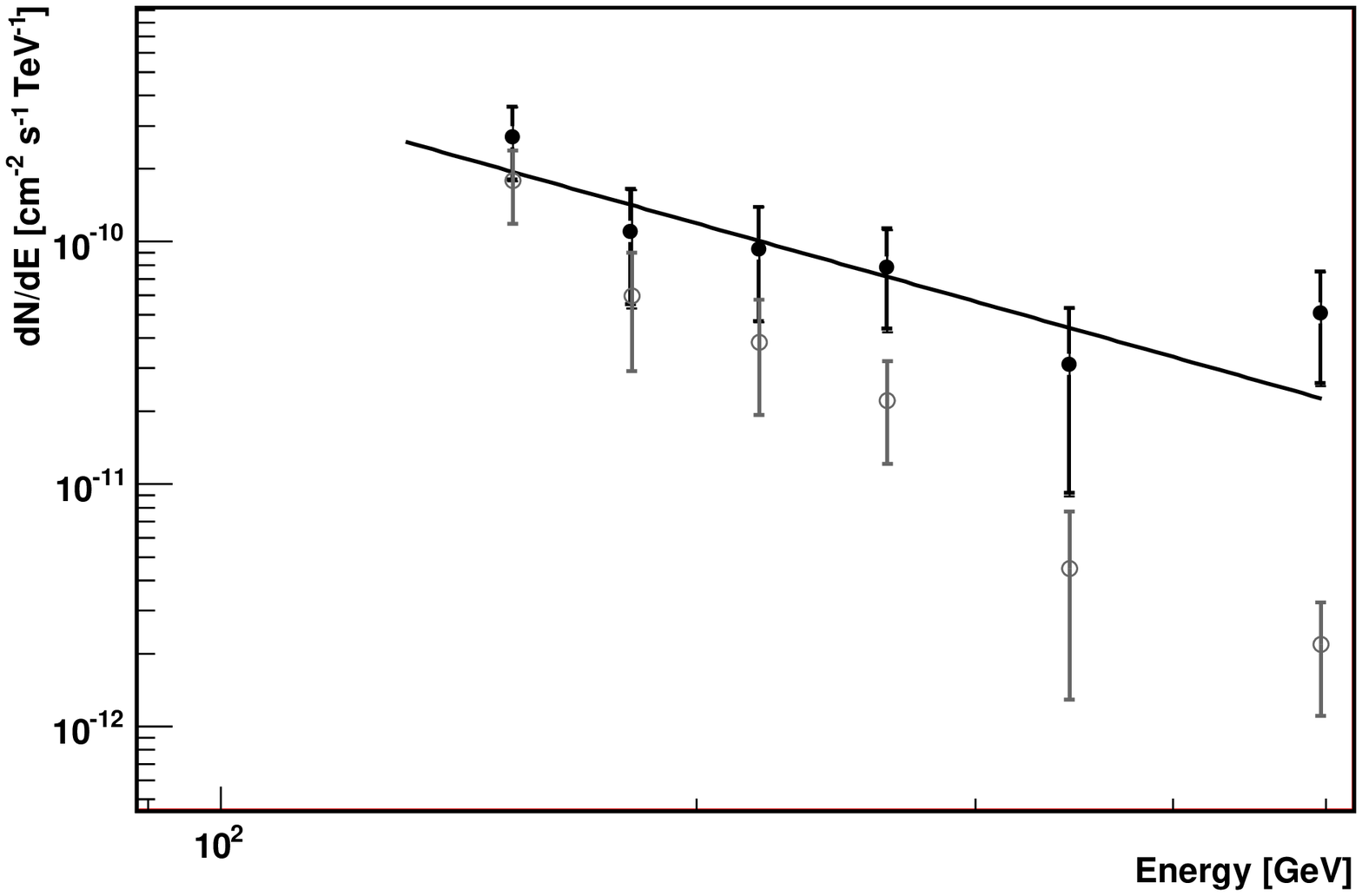}
     \caption{Left plot: power law index, $\Gamma$, of the deabsorbed spectrum of PKS~1424+240 as a function of the redshift, $z$, using Franceschini EBL model, with a 1\,$\sigma$ confidence belt. The vertical line marks the redshift, $z^*$~=~0.45, at which the slope of the deabsorbed spectrum equals the slope measured by {\it Fermi} at lower energies. Right plot:  measured (open points) and deabsorbed (filled points) spectrum of PKS~1424+240 at redshift $z$~=~0.45.}
     \label{1424_plot}
   \end{figure}

As a final application, we apply our method
to PKS~1424+240, a blazar of unknown redshift
recently observed in the VHE regime by VERITAS \cite{acciari10}. 
The slope  measured by {\it Fermi}/LAT in the energy range 0.1--100\,GeV is 
$1.83\pm0.03$. 
The corresponding $z^*$ at which the slope of the deabsorbed TeV spectrum 
becomes equal to it, is $0.45~\pm~ 0.15$,
using the Franceschini EBL model, see Fig.~\ref{1424_plot}.
This result is in agreement with the value of $0.5\pm0.1$, 
reported in \cite{acciari10}, calculated by applying the same procedure 
but using only simultaneous LAT data.
Our estimate on the most probable distance for PKS~1424+240
is 0.26~$\pm$~0.05, where the
error  is the $\sigma$ of the Gaussian 
fitting the $\Delta$z distribution. 

\end{document}